\documentclass[nofootinbib,pre,twocolumn,showpacs,preprintnumbers,amsmath,amssymb,superscriptaddress
]{revtex4-1}
\usepackage[normalem]{ulem}

\usepackage{graphicx}
\usepackage{color} 
\usepackage{amsmath}
\begin{document}-
\title{Specific heat and non-linear susceptibility in spin glasses with random fields}

 \author{M. V. Romitti} 
 \affiliation{PGFisica, Universidade Federal de Santa Maria, 97105-900 Santa Maria, RS, Brazil}%
 \author{F. M. Zimmer}\email{fabiozimmer@gmail.com}
 \affiliation{PGFisica, Universidade Federal de Santa Maria, 97105-900 Santa Maria, RS, Brazil}%
 \affiliation{Instituto de Fisica, Universidade Federal de Mato Grosso do Sul,  79070-900 Campo Grande, Brazil}
 \author{C. V. Morais}
 \affiliation{Instituto de Fisica e Matematica, Universidade Federal 
 de Pelotas, 96010-900  Pelotas, RS, Brazil}
\author{S. G. Magalhaes}\email{sgmagal@gmail.com}
\affiliation{Instituto de Fisica, Universidade Federal do Rio Grande do Sul, 91501-970 Porto Alegre, RS, Brazil}

\begin{abstract}
We study magnetic properties of spin glass SG systems under a random field (RF), beased on the suggestion that RFs can be induced by a weak transverse field in the compound LiHo$_x$Y$_{1-x}$F$_4$. 
We consider a cluster spin model that allows long-range disordered interactions among clusters and short-range interactions inside the clusters, besides a local RF for each spin following a Gaussian distribution with standard deviation $\Delta$.
We adopt the one-step replica symmetry breaking (RSB) approach to get an exactly solvable single-cluster problem. We discuss the behavior of order parameters, specific heat $C_\textnormal{m}$, nonlinear susceptibility $\chi_3$ and phase diagrams for different disorder configurations. In the absence of RF, the $\chi_3$ exhibits a divergence at $T_f$, while the $C_\textnormal{m}$ shows a broad maximum at a temperature $T^{**}$ around 30$\%$ above $T_f$, as expected for conventional SG systems. The presence of RF changes this scenario. The $C_\textnormal{m}$ still shows the maximum at $T^{**}$ that is weakly dependent on $\Delta$. However, the $T_f$ is displaced to lower temperatures, enhancing considerable the ration $T^{**}/T_f$. Furthermore, the divergence in $\chi_3$ is replaced by a rounded maximum at a temperature $T^{*}$, which becomes increasingly higher than $T_f$ as $\Delta$ enhances. As a consequence, the paramagnetic phase is unfolded in three regions: (i) a conventional paramagnetism ($T>T^{**}$; (ii) a region with formation of short-range order with frozen spins ($T^{*}<T<T^{**}$); (iii) a region with slow growth of free-energy barriers slowing down the spin dynamics before the SG transition ($T_f<T<T^{*}$) suggesting an intermediate Griffiths phase before the SG state. Our results reproduce qualitatively some findings of LiHo$_x$Y$_{1-x}$F$_4$ as the rounded maximum of $\chi_3$ behavior triggered by RF and the deviation of the conventional relationship between the $T_f$ and $T^{**}$.
\vskip \baselineskip

\end{abstract}
\maketitle

\section{Introduction}

There have been an intense experimental and theoretical debate about the  behavior of the non-linear susceptibility $\chi_3$ in the diluted Ising dipolar ferromagnetic 
compound LiHo$_x$Y$_{1-x}$F$_4$ (see, for instance, \cite{jonsson2007, ancona2008}). 
This quantity is considered one of the fingerprints of a second order spin glass (SG) transition \cite{Binder86}.
In the absence of an applied transverse field $H_t$, the SG transition in this compound is well captured experimentally by $\chi_3$  that diverges at the so called freezing temperature $T_f^{0}$ \cite{Wu91}. 
Thus, one could expect that the application of $H_t$  would lead  LiHo$_x$Y$_{1-x}$F$_4$ to behave as a quantum Ising SG  \cite{Mydosh2015,Gingras11}. However, it was  found that the divergence in $\chi_3$ is smoothed out and replaced by a rounded maximum at a certain temperature $T^*$ lower than $T_f^{0}$. A possible explanation  considers  random  fields (RFs) induced by the coupling between $H_t$ with the off diagonal terms of the dipolar interactions in  LiHo$_x$Y$_{1-x}$F$_4$ \cite{Laflorencie2006,Gingras2006}. This mechanism allows to reproduce successfully the experimental behavior of $\chi_3$ \cite{Gingras2006,Magalhaes2017}. 

 For instance, in the LiHo$_{x}$Y$_{1-x}$F$_4$ for $H_t =0$, the single ion ground state is a doublet (or equivalently up and down Ising spin states) separated from the first excited state by an energy $\sim 9.4$ K  \cite{Wu991}.  For the temperature range of  interest, only these two Ising state are significantly populated. This degeneracy is lifted when $H_t$ is turned on, carrying to quantum mechanical mixing of the up and down states. The microscopic hamiltonian of this compound considers basically dipolar interactions \cite{Gingras2006} that can be projected into these two states resulting in an effective disordered Ising model with an effective transverse field $\Gamma$ proportional to the energy splitting between the up and down Ising states \cite{Chakraborty2004}. Moreover, the off-diagonal terms of the dipolar interactions are not canceled by symmetry in the disordered compound ($x<1$), generating an effective longitudinal RF which is dependent of $H_t$.  
%
%
For small intensities of $H_t$ (for instance, $H_t < 0.5 T$ ), the energy splitting   of the Ising ground state doublet is negligible\footnote{ This situation refers to the paramagnetic phase.}. Thus, the quantum tunneling between the up and down Ising states can be also neglected \cite{Wu991}.
%
This scenario corresponds to a semi-classical regime. 
On the other hand, there are experimental results indicating that even for this regime, RF can be already quite active.  For instance,  Ref. \cite{jonsson2007} displays not only $\chi_3$ vs $H_t$ but also $T^*$ vs $H_t$ for LiHo$_{0.165}$Y$_{0.835}$F$_{4}$. In the first case, the divergence of $\chi_3$ is clearly replaced by a maximum already for 
$H_t < 0.2 T$.  In the same range,  $T^*$ is 
clearly decreasing as $H_t$ is increased.

Very recently,
 the $\chi_3$ has been theoretically studied using the induced RF mechanism  in the quantum \cite{Magalhaes2017} and  semi-classical regimes \cite{Morais2016}. In this last regime,  as pointed above, it is considered a weak $H_t$, enough to induce a RF but not enough to lead the quantum fluctuations dominate the thermal ones. 
Thereby, the conceptual and mathematical framework of the classical replica  mean field theory for SG  can be applied \cite{SK,Parisi80a,Parisi80b,SNA}. 
In that framework, it is known that in the absence of RF,  $\chi_3$ can be directly 
related
with the inverse of the eigenvalue replicon $\lambda_\textnormal{AT}$ \cite{Binder86}. Thus, $\chi_3$ diverges 
exactly as  $\lambda_\textnormal{AT}= 0$ which 
is the onset of the replica symmetry breaking (RSB) SG phase at the freezing temperature $T_f^0$. However, results from Ref. \cite{Morais2016} showed that as RF is present the situation is more complex,  the relationship of $\chi_3$ with $\lambda_\textnormal{AT}$ is changed (see also Refs. \cite{Pirc1987,Kopec1990}). Nevertheless,
$\lambda_\textnormal{AT}=0$ still locates the RSB SG transition at $T_f$ (the freezing temperature with RF).
Therefore, the divergence found in $\chi_3$ is replaced by a rounded maximum  located at $T^*$ which no longer coincides with $T_f$. In fact, $T^*$ is increasingly higher than $T_f$  but decreases as the RF effects become more important \cite{Morais2016}.  As a consequence, the paramagnetic phase is unfolded in two regions, with one of them probably being a Griffiths phase \cite{Biltmo2012}.
It should be remarked that this phase diagram is basically preserved in the quantum regime \cite{Magalhaes2017}.

Another probe which can bring information 
on the LiHo$_x$Y$_{1-x}$F$_4$ physics is the magnetic specific heat $C_\textnormal{m}$. 
It is well known from  
conventional SG systems that the experimental
$C_\textnormal{m}$ does not present any  sharp anomaly at the SG transition  being the critical exponent  $\alpha$ negative, around $-2$ (see Ref. \cite{Hertz} and references therein). Actually, $C_\textnormal{m}$ presents a broad maximum around a certain temperature $T^{**}$ estimated to be $20 \%$-$40\%$ higher than the freezing temperature \cite{Binder86}.This broad maximum in the $C_\textnormal{m}$ located above the SG transition temperature is one of  the most important experimental fingerprints of the usual SG behaviour. Interestingly, as for $\chi_3$, the experimental behavior of specific heat in the LiHo$_x$Y$_{1-x}$F$_4$
also has controversies even in absence of $H_t$. In the extreme dilution limit $x=0.045$, the debate focused on the question whether exists a spin liquid antiglass as ground state  instead of a SG-like state  \cite{Gosh03}. 
The antiglass scenario is based on the observation of sharp peaks in the specific heat.  In contrast, Quilliam and collaborators \cite{Quilliam07}  
observed a  broad maximum for the specific heat, as expected for a SG-like state. 
This debate is a quite clear indication that specific heat is also source  of interesting information on the complex physics of the  LiHo$_x$Y$_{1-x}$F$_4$ \footnote{There is a clear disagreement for $C_\textnormal{m}$ between the experiment 
and the classical Monte Carlo results \cite{Biltmo2012} even in zero $H_t$. }. However, not much attention has been given to the behavior of magnetic specific heat in the presence of $H_t$. Such kind of study can be helpful to complement and  clarify the puzzling situation presented by $\chi_3$ behavior described above.

The goal of the present work is to provide an unified  analytical description of the behavior of magnetic specific heat and $\chi_3$ of an SG model with the presence of a RF. It is assumed that even weak $H_t$ can
induce a RF as proposed by Refs. \cite{Laflorencie2006,Gingras2006}.  Firstly, it should be emphasized that the Sherrington-Kirkpatrick (SK) theory  \cite{SK},  which is the standard mean-field procedure for SG was very successful in explaining many aspects of the  experimental behavior of spin glass systems, except, the $C_\textnormal{m}$. Actually, this approach  predicts the $C_\textnormal{m}$ with a sharp cusp at the freezing temperature.
This result is not consistent with the observed behaviour of $C_\textnormal{m}$.
In order to overcome this flaw of the SK theory, we adopt the cluster formulation proposed by Soukoulis and Levin \cite{Soukoulis781,Soukoulis782}. 
In that proposal, starting from the Ising model,
cluster of spins are used not only to provide intraclusters short range spin correlations necessary to fit the  experimental behavior of $C_\textnormal{m}$ but also to stabilize the SG state. It means that is possible to obtain a $C_\textnormal{m}$ curve with a broad maximum at a temperature above $T_f$.
In the effective SG model which results from the cluster formulation, the quenched bond disorder appears as an intercluster interaction.
Actually, one has two coupled problems. The first one refers to the intercluster disordered interaction which
is solved exactly at mean field replica method since it is considered infinite ranged intercluster interaction.   
The second one refers to the intracluster problem. 
For this part, one should consider clusters with a certain inner structure.  In fact, the intercluster contribution exhibits a cusp at $T_f$. However, as long as the intracluster contribution becomes dominant (by increasing the cluster size) the cusp tends to disappear and it appears  a broad maximum  located at a certain temperature which is higher than  $T_f$ in qualitative agreement with the experimental behaviour of Cm.  
Specifically for our case, it is used a cubic intracluster structure considering that an uniform ferromagnetic interaction is present 
between nearest neighbours and the RF acts in each spin. As a result,  the intracluster Ising spin degrees of freedom of a finite cluster are computed by exact enumeration for each RF configuration. 
This  crucial step allows, then, to go back to the intercluster problem, i. e. the SG problem, which is treated within the one-step replica symmetry breaking (RSB) scheme.  In our work,  there is no RSB without random bonds.

We  highlight that in this cluster mean field theory for SG, we investigated in detail the roles of the intracluster and the intercluster  parts to determine the behavior not only of $C_\textnormal{m}$ but also of $\chi_3$. 
As discussed previously, $\chi_3$ has been in the center of intense debate which, ultimately, deals with the existence of the SG state in the LiHo$_x$Y$_{1-x}$F$_4$.  
In particular, there are also issues concerning the presence of SG in uniform external magnetic field $h$. For instance, simulations on three-dimensional Ising SG model have pointed inconclusive results concerning the existence of SG state in presence of $h$ \cite{Baity2014, Baity2014Jstat}, while it is well established that mean-field studies found SG state by means of the Almeida and Thouless analysis ($\lambda_\textnormal{AT}$), in the so called RSB picture. In the present study, we shall demonstrate that it is still preserved the relationship  $\chi_3 \sim \lambda_\textnormal{AT}^{-1}$ in the cluster formulation without RF.
In the presence of RF, that relationship is modified.
As a consequence, one can expect that shall emerge three energy scales:
(i) the RSB freezing temperature $T_f$; (ii)  $T^*$ associated with the rounded maximum of $\chi_3$  and  (iii) $T^{**}$ associated with the broad maximum of $C_\textnormal{m}$. It should be noticed that $T^*$ would exist only for finite RF. Actually, the central question of the present work is how $T_f$, $T^*$ and $T^{**}$ evolve as the RF effects are enhanced.
Indeed, the behavior of $T^*$ and $T^{**}$ would indicate how the paramagnetic (PM) phase is unfolded in PM sub-regions displaying distinct spin correlations.

This paper is structured as follows. In Sec. \ref{model} we discuss the model and the analytic calculations to get  the order parameters  and thermodynamic quantities as $\chi_3$ and $C_v$. Our numerical results are presented in Sec. \ref{res}. In Sec. \ref{conc} we present the conclusion.

\section{Model}\label{model}

We follow closely the cluster SG mean field theory proposed in Ref. \cite{Soukoulis781}  rewriting the Ising model with RFs, $ H = - \sum_{i,j}^{N} J_{ij} \sigma_i \sigma_j - \sum_{i=1}^{N} h_{i} \sigma_{i}$, in terms of $N_{cl}$ spin clusters with $n_s$ spins inside of the $\nu$-th cluster, hence  $N=N_{cl}n_s$. 
This procedure defines a new  variable $\sigma_{\nu}=\sum_{i}^{n_s} \sigma_{\nu,i}$ and the effective hamiltonian becomes
\begin{equation}
H=-\sum_{\nu<\lambda}J_{\nu\lambda}\sigma_{\nu}\sigma_{\lambda}-\sum_{\nu}(\sum_{i<j}J_1\sigma_{\nu,i}\sigma_{\nu,j}+\sum_{i}h_{\nu,i}\sigma_{\nu
,i})
\label{ham}.
\end{equation}
At this stage, the only approximation 
is to consider that the clusters are separated by a distance far larger than their average size. Thus, the intercluster interaction is assumed to be independent of the site position inside the cluster \cite{Soukoulis782}.
It means that the neighboring clusters interact only via exchange interactions $J_{\nu\lambda}$ between total spins on each cluster with intracluster short-range interactions $J_1$. 
In the model (\ref{ham}), the two sources of disorders $J_{\nu\lambda}$ and random fields $h_{i}$ follow the probability distributions given below
\begin{equation}
P(J_{\nu\lambda})=\frac{1}{\sqrt{2\pi \frac{J^2}{N_{cl}}}}\exp{\left[-\frac{1}{2}\left( { \frac{J_{\nu\lambda}}{\frac{J}{\sqrt{N_{cl}}}}}\right)^2\right]}
\end{equation}
and
\begin{equation}
\overline{P}(h_{\nu_i})=\frac{1}{\sqrt{2\pi \Delta^2}}\exp{\left[-\frac{1}{2}\left( { \frac{h_{\nu_i}}{\Delta}}\right)^2\right]}.
\end{equation}

We use the same procedure as in Ref. \cite{SNA} to obtain the free energy per cluster
$f=-1/(\beta N_{cl}) \langle \langle \mbox{ln} Z(\left\{J_{ij}\right\},\left\{h_i\right\}) \rangle \rangle_{J,h}$, where $Z(\left\{J_{ij}\right\},\left\{h_i\right\})$ is the partition function for a given quenched distribution of the random couplings and fields. $\langle \langle  . . \rangle \rangle_{J,h}$ denotes averages over these disorders and $\beta=1/T$.
As usual, the replica method is applied in order to calculate the quenched disorders:
\begin{equation}
-\beta f=
\lim_{n\rightarrow 0}\frac{1}{N_{cl}n}
\left(\langle \langle Z(\left\{J_{ij}\right\},\left\{h_i\right\})^{n} \rangle \rangle_{J,h}-1\right).
\label{eq4}
\end{equation}
The average over the random couplings can be evaluated and the replicated partition function for a given distribution of RF 
is expressed as
\begin{equation}
\begin{split}
\langle\langle Z(\left\{h_i\right\})^{n}\rangle\rangle = 
\left\langle \text{Tr} \exp\left[-\beta\sum_{\nu}\sum_{\alpha}H_\textnormal{intra}^{\alpha}(\nu,\{h_i\}) \right. \right. \\ 
\left.\left.+\frac{\beta ^2 J^2}{4N_{cl}}\sum_{\alpha,\gamma}\left( \sum_{\nu} \sigma_{\nu}^{\alpha} \sigma_{\nu}^{\gamma}\right)^2\right]\right\rangle_{h_i},
\end{split}
\end{equation}
where Tr is the trace over spin variable, $\alpha$ and $\gamma$ are the replica indices with the $\sum_{\alpha,\gamma}$ considering $\alpha$ and $\gamma=1,\cdots n$, $\langle\cdots\rangle_{h_i}$ denotes the average over the RF distribution, and 
$H_\textnormal{intra}^{\alpha}(\nu,\{h_i\})=-\sum_{i<j}J_1\sigma^{\alpha}_{\nu,i}\sigma^{\alpha}_{\nu,j}-\sum_{i}h_{\nu,i}\sigma^{\alpha}_{\nu
,i}$ is represents the intracluster terms.
The quadratic terms are linearized by introducing the SG order parameters:
\begin{equation}
\begin{split}
\langle Z(\{h_i\})^{n}\rangle = \int Dq_{\alpha,\gamma} \exp \left\lbrace -N_{cl}\left[\frac{\beta ^2 J^2}{4}\sum_{\alpha,\gamma}q_{\alpha,\gamma}^2 \right.\right. \\
\left.\left.-\frac{1}{N_{cl}}\left\langle \ln \mbox{Tr}\exp{[-\beta\sum_{\nu} H_\textnormal{eff}(\nu,\{h_i\})]} \right\rangle_{h_i} \right] \right\rbrace
\end{split}
\end{equation}
where
\begin{equation}
H_\textnormal{eff}(\nu,\{h_i\})=\sum_{\alpha}H_\textnormal{intra}^{\alpha}(\nu,\{h_i\}) -\frac{\beta J^2}{2}\sum_{\alpha,\gamma}q_{\alpha\gamma}\sigma_{\nu}^{\alpha}\sigma_{\nu}^{\gamma}
\end{equation}
is an effective single-cluster model with interacting replicas.
In the thermodynamic limit, the functional integrals over $q_{\alpha\gamma}$ are obtained from the saddle-point method:
\begin{equation}
q_{\alpha\gamma}=\langle\frac{\mbox{Tr} \sigma^{\alpha}\sigma^{\gamma}\exp[-\beta H_\textnormal{eff}(\{h_i\})}
{\mbox{Tr} \exp[-\beta H_\textnormal{eff}(\{h_i\})}\rangle_{h_i},
\end{equation}
in which this correlation for the same replica ($\alpha=\gamma$) can be associated with the expectation value of the cluster magnetic moment magnitude, while for different replicas ($\alpha\ne\gamma$) it is related to the SG order parameter.

The one-step replica symmetry breaking 1S-RSB is used to parametrize the replica matrix as: $\bar{q}=q_{\alpha\alpha}$, $q_0=q_{\alpha\gamma}$ if $I(\alpha/x)=I(\gamma/x)$ or $q_1=q_{\alpha\gamma}$ if $I(\alpha/x)\neq I(\gamma/x)$, where $I(y)$ is the smallest integer greater than $y$ and $x$ is the size of diagonal blocks of the replica matrix with 1S-RSB solution \cite{Parisi80a}. This anzatz results in the following free energy expression
\begin{equation}\begin{split}
\beta f = \frac{J^2\beta^2}{4} (\bar{q}^2+ x (q_1^2-q_0^2)-q_1^2)- \\
-\frac{1}{x} \left\langle\int Dz \mbox{ln} \int Dv [K(\{h_i\},v,z)]^x\right\rangle_{h_i}
\end{split}\label{freeenergy}\end{equation}
with $\int D\xi=\int d\xi \mbox{e}^{-\xi^2/2}/\sqrt{2\pi}$ ($\xi=z$ or $v$),
\begin{equation}
K(\{h_i\},v,z) = \mbox{Tr}\exp{[-\beta H_\textnormal{eff}^\textnormal{1S}(\{h_i\},v,z)]}
\end{equation}
and the effective single-cluster model
\begin{equation}
\begin{split}
H_\textnormal{eff}^\textnormal{1S}(\{h_i\},v,z) =-\left(J\sqrt{q_1-q_0}v+J\sqrt{q_0}z\right)\sigma \\- \frac{\beta J^2}{2}(\overline{q}-q_1)\sigma^2  -\sum_{i,j}^{n_s}J_{1}\sigma_{i}\sigma_{j}-\sum_{i}^{n_s}h_{i}\sigma_{i},
\end{split}\label{heffective}
\end{equation}
where the order parameters are exhibited in appendix \ref{appendixa}.

Other thermodynamic quantities can now be obtained from the free energy. For instance, the linear susceptibility $\chi_1$ is given by $\chi_1 = \beta[\bar{q} - q_1+ x(q_1-q_0)]$~\cite{Parisi80b}. 
The nonlinear susceptibility $\chi_3$ can be derivated from $\chi_3= -\frac{1}{3!} \frac{ \partial^2 \chi_1}{ \partial h^2} |_{h \rightarrow 0 }$, where $h$ is an applied longitudinal magnetic field. 
The internal energy ( $u=-\frac{\partial}{\partial \beta}(\beta f)$) and the specific heat ($C_\textnormal{m}=\frac{d}{dT} u$)  are also obtained:
\begin{equation}
C_\textnormal{m}= C_\textnormal{inter}+C_\textnormal{intra}+C_\textnormal{RF}
\end{equation}
where
\begin{equation}
C_\textnormal{inter}=\frac{d}{dT}[\frac{\beta J^2}{2} (\bar{q}^2+ x(q_1^2-q_0^2)-q_1^2)]\label{cm_inter},\end{equation}
\begin{equation}C_\textnormal{intra}=J_1\frac{d}{dT}\langle \int Dz \frac{\int Dv K^{x-1}\mbox{Tr}\sum_{(i,j)}^{n_s}\sigma_{i}\sigma_{j}\mbox{e}^{-\beta H_\textnormal{eff}^\textnormal{1S}}}{\int Dv K^{x}}\rangle_{h_i},
\label{cm_intra}\end{equation}
\begin{equation}
C_\textnormal{RF}=\frac{d}{dT}\langle \int Dz \frac{\int Dv K^{x-1}\mbox{Tr}\sum_{i}^{n_s}h_{i}\sigma_{i}\mbox{e}^{-\beta H_\textnormal{eff}^\textnormal{1S}}}{\int Dv K^{x}}\rangle_{h_i}.
\label{cm_RF}\end{equation}

In particular, the replica symmetry solution can occur at high temperatures when $q=q_0=q_1$, resulting in the
following effective model
\begin{equation}\begin{split}
 H_\textnormal{eff}^\textnormal{RS}(\{h_i\}) =-J\sqrt{q}z\sigma - \frac{\beta J^2}{2}(\overline{q}-q)\sigma^2 -H_\textnormal{intra}(\{h_i\}).
\end{split}\label{hrs}
\end{equation}
The stability of the RS solution can be obtained from the de Almeida-Thouless eigenvalue \cite{Almeida78} given by Eq. (\ref{at}).

The nonlinear susceptibility $\chi_3$ is obtained within the RS solution from:
\begin{equation}
 \chi_3=\left.\frac{1}{3!}\frac{d^3 m(q,\bar{q},h)}{d h^3}\right|_{h=0},
 \label{C1}
 \end{equation}
 where 
 \begin{equation}
  m(q,\bar{q},h)= \left\langle \int Dz \frac{\mbox{Tr} ~\sigma \exp{(-\beta H_\textnormal{eff}^\textnormal{RS}(\{h_i\}))}}{\mbox{Tr} \exp{(-\beta H_\textnormal{eff}^\textnormal{RS}(\{h_i\}))}}\right\rangle_{h
_i}
 \label{C2}
 \end{equation}

In Appendix \ref{appendixc}, we develop 
an explicitly form for $\chi_3$ in terms of spin correlations which is given in Eq. (\ref{C8}). 
This form 
can also be expressed 
directly in terms of $\lambda_\textnormal{AT}$ (see Appendix \ref{appendixb}) as
\begin{equation}
 \chi_3=\frac{\beta^3}{3} \left [\frac{3}{b(\lambda_\textnormal{AT})}-1\right]V_2
 \label{eq_bat}
\end{equation}
where $b(\lambda_\textnormal{AT})=(2-\beta^2J^2 V_3)(\lambda_\textnormal{AT} - V_5)- \beta^4 J^4 V_2 V_4$ with 
\begin{equation}\begin{split}
 V_2=\langle\int Dz[ \langle \sigma^4\rangle - 4 \langle \sigma\rangle \langle \sigma^3\rangle
-3 \langle \sigma^2\rangle^2 
\\ + 12\langle \sigma\rangle^2\langle \sigma^2\rangle - 6\langle \sigma\rangle^4 ]\rangle_{h_i},
 \end{split}
 \label{V2}
 \end{equation}
 \begin{equation}\begin{split}
 V_3=\langle\int Dz[ \langle \sigma^4\rangle - 2 \langle\sigma\rangle \langle\sigma^3\rangle
-\langle \sigma^2\rangle^2 
+ 2\langle\sigma\rangle^2\langle\sigma^2\rangle]\rangle_{h_i} ,
 \end{split}
 \label{V3}
 \end{equation} 
\begin{equation}\begin{split}
 V_4=\left\langle\int Dz[\langle\sigma\rangle \langle\sigma^3\rangle
-\langle\sigma\rangle^2\langle\sigma^2\rangle]\right\rangle_{h_i}
\end{split}
\label{V4}
\end{equation}
and 
\begin{equation}\begin{split}
 V_5=\beta^2 J^2\langle\int Dz[ \langle\sigma\rangle \langle\sigma^3\rangle -3 \langle \sigma\rangle^2\langle\sigma^2\rangle
+2\langle \sigma\rangle^4\rangle]\rangle_{h_i},
 \end{split}
 \label{V5}
 \end{equation} 
in which $\langle \cdots\rangle$ represents the thermal average over the effective RS model $H_\textnormal{eff}^\textnormal{RS}$ with $h=0$.
Particularly, in absence of RF and for $T\geq T_f$ ($q=0$ with RS stable), we obtain $V_4=0$ and  $V_5=0$,  resulting in
\begin{equation}
 \chi_3(\Delta=0)=\frac{\beta^3}{3} \left[\frac{3}{(2-\beta^2 J^2 V_3)\lambda_\textnormal{AT}}  -1 \right]  V_2
\end{equation} 
Therefore, the $\chi_3$ diverges at $T_f$ in which $\lambda_\textnormal{AT}=0$.

\begin{figure}[ht]
\center{\includegraphics[angle=00,width=\columnwidth]{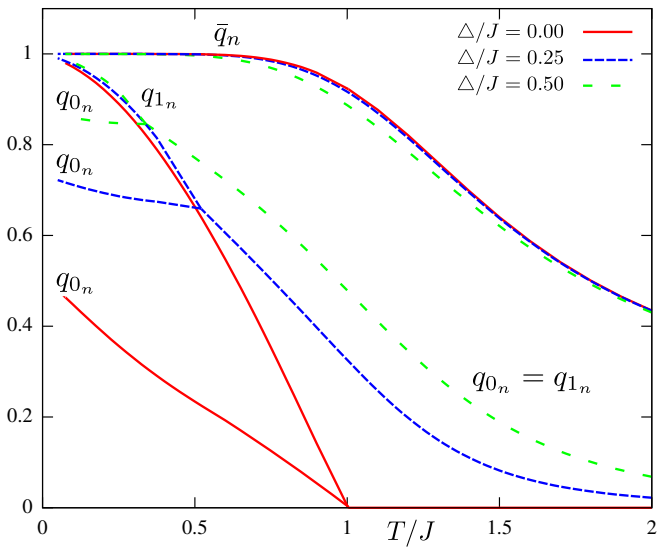}}
\caption{Normalized one step RSB order parameters as a function of the temperature for several values of $\Delta$ when $J_1/J=0.70$ and eight spins per cluster $n_s=8$ assuming a simple cubic lattice shape. 
Here, $q_{0_n}=q_0/n_s^2$, $q_{1_n}=q_1/n_s^2$ and $\bar{q}_{n}=\bar{q}/n_s^2$. }  
\label{fig_op}  \end{figure}

\section{Results}\label{res}
In this section, we present numerical results obtained from the single-cluster problem (Eqs. (\ref{freeenergy})-(\ref{heffective})). The behavior of the SG order parameters, linear $\chi_1$ and  nonlinear $\chi_3$ susceptibilities, and specific heat $C_\textnormal{m}$ are analyzed 
considering combined variations of the parameters $\Delta/J$, $J_1/J$ and $T/J$ for clusters following a simple cubic lattice shape with 8 spins.
In particular, the 1S-RSB and the replica symmetry stability (de Almeida-Thouless (AT) line) are used in order to locate the freezing temperature $T_f$. 
\begin{figure}[t]
\center{
\includegraphics[width=1.0\columnwidth]{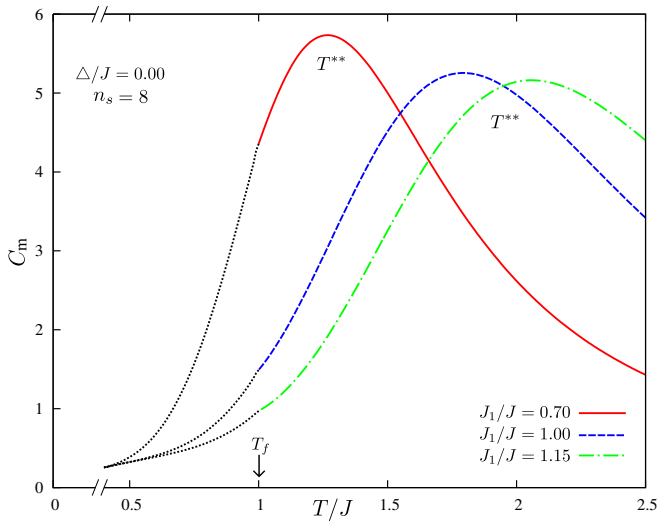}} 
\caption{Specific heat as a function of temperature for different ferromagnetic intracluster interactions without the presence of  RF. 
The dotted lines represent results of unstable RS solution ($\lambda_\textnormal{AT}<0$). The $C_\textnormal{m}$ values located below the axis break are not significant and numerically reliable.}
\label{fig1aa}  \end{figure}

\begin{figure}[t]
\center{\includegraphics[width=1.\columnwidth]{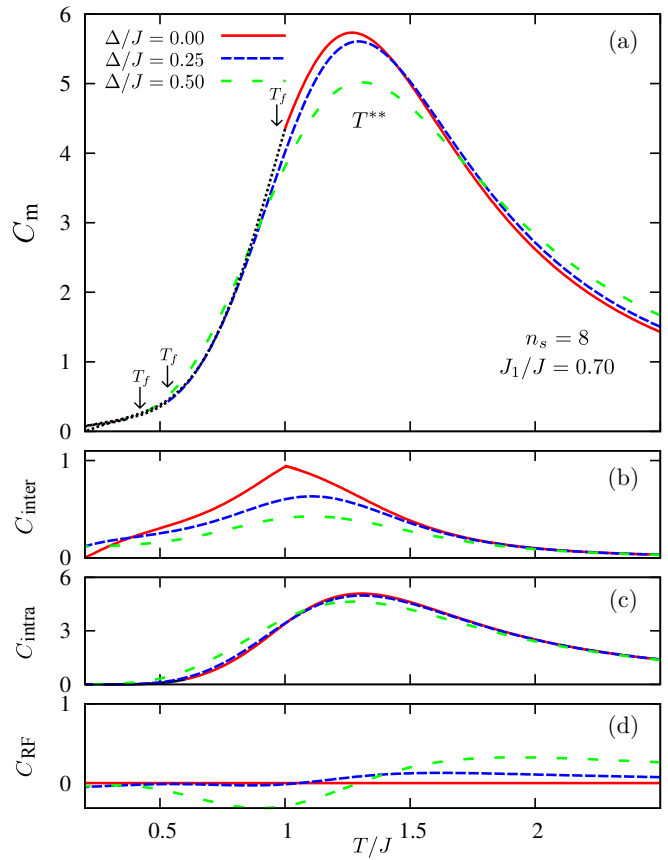}} 
\caption{ (a) $C_\textnormal{m}$ versus $T/J$ different values of $\Delta$ ($0.00, 0.25J$ and $0.50J$) when $J_1/J=0.70$ is keeping constant for simple cubic lattice clusters with 8 spins. Panels (b), (c) and (d) show the contributions from intercluster and intracluster interactions, and the explicit RF effects for the specific heat, respectively (see Eqs. (\ref{cm_inter}), (\ref{cm_intra}) and (\ref{cm_RF})).  }  
\label{fig_Cm_RF}  
\end{figure}

Fig. \ref{fig_op} shows the 1S-RSB SG order parameters as a function of the temperature. For instance, $q_0$ and $q_1$ exhibit a transition from the RS behavior ($q=q_0=q_1$) to a RSB region ($q_0\ne q_1$) at the freezing temperature $T_f$. The RFs induce these order parameters even in the RS region. In addition, the RFs displace the $T_f$ to lower temperatures. 
In particular, the transition from RS to RSB solution can also be located by the Almeida-Thouless line that  coincides with the beginning of the RSB. 
Furthermore, in the present cluster formalism,  the replica diagonal elements have an essential role. They are represented by 
$\bar{q}=\langle \sigma^{2}_{\nu} \rangle_{H_\textnormal{eff}}$ that can be interpreted as the intensity of the cluster magnetic moment \cite{Soukoulis781}. That is an important difference with Ref. \cite{Morais2016}. There, using the SK model, 
$\bar{q}=1$. Here, $\bar{q}$ depends on the temperature as well as the intracluster interactions and RFs.

The specific heat and susceptibilities are now analyzed in order to understand the effects of RFs on this SG problem. 
For instance, Fig. \ref{fig1aa} exhibits the $C_\textnormal{m}$ as a function of $T/J$ for different intracluster interactions in absence of RF ($\Delta=0$). 
The $C_\textnormal{m}$ curve presents a broad maximum at a temperature $T^{**}$ that depends on the intensity of $J_1$ (see Fig. \ref{fig1aa}). 
The increase of $J_1$ displaces $T^{**}$ to higher temperatures at the same time that the $C_\textnormal{m}$ maximum becomes lower. It means that the intracluster short-range interactions affect the specific heat behavior. 
At $T_f$ the $C_\textnormal{m}$ presents a small mark that is associated with the intercluster interactions. Specifically, this comes from the temperature derivative of the SG  order parameters (see Eq. \ref{cm_inter}) that become different from zero at $T_f$ (see Fig. \ref{fig_op} for $\Delta=0$).
It is also important to note that $T_f/J$ is keeping at unity and the maximum appears in a range of stable RS solution ($\lambda_\textnormal{AT}>0$). 
In other words, $T_f< T^{**}$ and the ratio $T^{**}/T_f$ can be adjusted by $J_1/J$ in order to get the behavior observed in canonical SG systems ($T^{**}/T_f\approx 1.30$), as instance $J_1/J=0.70$.

However, the presence of RFs changes this scenario. 
As shown in Fig. \ref{fig_Cm_RF}(a), the $C_\textnormal{m}$ still exhibits the broad maximum at $T^{**}$ that is weakly dependent on $\Delta$, but $T_f$ is decreased by RFs. 
As a consequence, 
$T^{**}/T_f$ grows when the RFs are considered.
The different contributions for $C_\textnormal{m}$ can be analyzed in   
 Figs. \ref{fig_Cm_RF}(b)-\ref{fig_Cm_RF}(d). For instance, the intercluster interaction contributions displayed in panel (b) indicate that the peak at $T_f$ vanishes in the presence of RFs. This occurs because the RF induces the SG order parameters at the whole range of temperature, avoiding the discontinuity in the derivative of these order parameters as discussed before. As a consequence, the $C_\textnormal{m}$ curve becomes smooth at $T_f$ (see Fig. \ref{fig_Cm_RF}(b)). 
 From the Fig. \ref{fig_Cm_RF}(c), one can see that the short-range intracluster represents the main contribution for $C_\textnormal{m}$. This contribution is weakly affected by the RF, at least in the range of low strength of RF adopted here. This explains why the $T^{**}$ position is slightly dependent on the RFs. Besides, the explicit RF contribution (Fig. \ref{fig_Cm_RF}(d)) has a lower intensity as compared with the intracluster one.     

\begin{figure}[htb]
\center{
\includegraphics[width=\columnwidth]{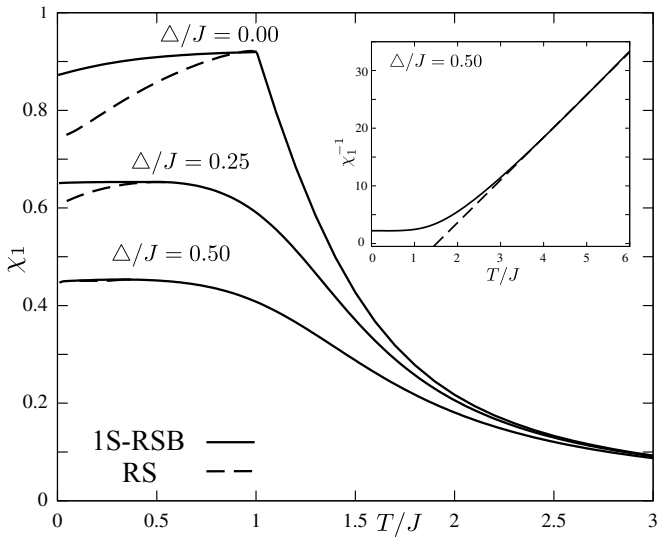}}
\caption{Linear susceptibility $\chi_1$ vs $T/J$ for $J_1/J=0.70$ and $n_s=8$ with $\Delta/J=0.00, 0.25$ and $0.50$. The solid and dashed lines correspond to 1S-RSB and RS solutions respectively.  
The detail presents $\chi_1^{-1}$ as a function of $T/J$ (full line) when $\Delta/J=0.50$. The dashed line represents the Curie-Weiss law linear extrapolation from higher temperatures. } \label{linear} 
\end{figure}

The linear susceptibility can be analyzed in Fig. \ref{linear}. 
For the absence of RF, $\chi_1$ presents a cusp at the freezing temperature, in which the $\chi_1$ becomes weakly dependent on the temperature within the RSB region, appearing a divergence between the results obtained with RS and 1S-RSB solutions. 
However, this cusp is suppressed in the presence of RF, but the divergence between both solutions is still present with a weak dependence on the temperature for the 1S-RSB.    
In addition, the detail of Fig. \ref{linear} exhibits the reciprocal of $\chi_1$ that follows a Curie-Weiss behavior at higher temperatures ($T/J\gtrsim 3T_f/J$).

\begin{figure}[ht]
\center{\includegraphics[width=\columnwidth]{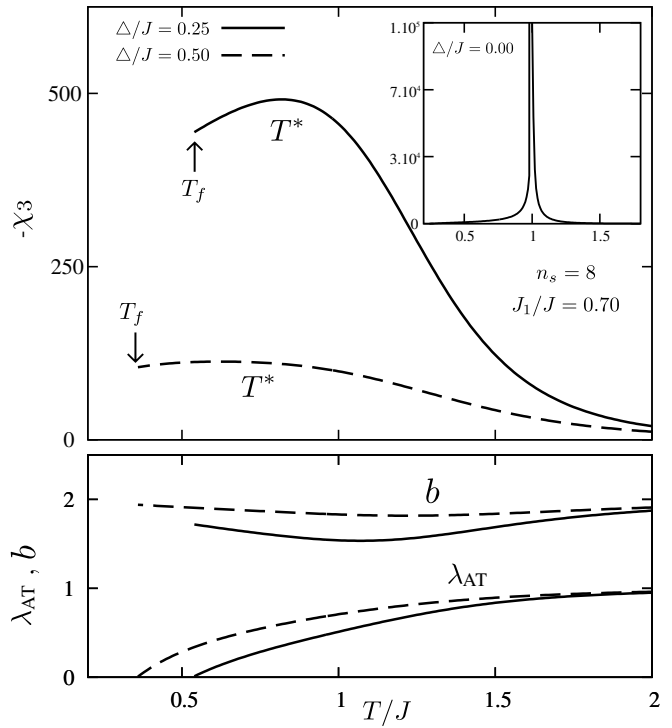}} 
\caption{(a) Nonlinear susceptibility $\chi$ vs $T/J$ for intracluster interaction $J_1/J=0.70$ with $\Delta/J=0.25$ and 0.5. The arrows locate the limit of RS stable solution. The inset shows the  $\chi_3$ vs $T/J$ for $\Delta/J=0.00$. Panel (b) presents the behavior of $\lambda_\textnormal{AT}$ and the denominator $b$ of the $\chi_3$ expression for $\Delta>0$.  }  
\label{fig1a} 
\end{figure}

\begin{figure}[ht]
\center{
\includegraphics[width=\columnwidth]{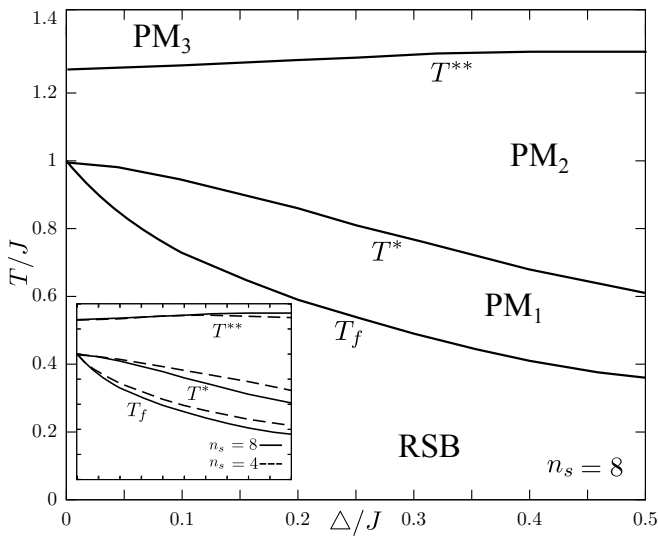}}
\caption{
Phase diagram $T/J$ vs $\Delta$ for $n_s = 8$ and $J_1/J = 0.70$ that shows $T_f$ separating the RSB solution (an SG state) from the RS PM phases. The figure also displays the crossover temperatures $T^{**}$ and $T^*$ delimiting the different PM behaviors (PM$_1$, PM$_2$ and PM$_3$) associated with $C_\textnormal{m}$ and $\chi_3$ maximum respectively. The inset exhibits a comparison between results obtained from $n_s=4$  and $n_s=8$, where no significant changes are observed.}  
\label{fig2a} 
\end{figure}

Another relevant result can be derived from the higher order susceptibility terms that are more sensitive to the SG phase transition. For instance, Fig. \ref{fig1a}(a) displays the nonlinear susceptibility $\chi_3$ as a function of $T/J$ for clusters with $n_s=8$ and $J_1= 0.70$ when different values of $\Delta$ are considered. 
The $\chi_3$ result for $\Delta = 0$ shows a divergence at the freezing temperature $T_f$ (see inset of Fig. \ref{fig1a}(a)), identifying the SG phase transition in absence of RF. 
However, this divergent peak becomes a rounded maximum at a temperature $T^{*}$ as $\Delta$ increases.
In particular, $T^*$ does not match anymore with the transition temperature $T_f$ 
as can be seen Fig. \ref{fig1a}(b) from the $\lambda_\textnormal{AT}$ curve. 
It is important to remark that, in this case, $T^*$ occurs in a range of temperature where the RS solution is stable, i.e, $T_f < T^{*}$ for a given $\Delta$.
It also means that the temperature indicated by $T^*$ in the presence of RF does not locate an SG phase transition. 
This $T^*$ displacing from $T_f$ can be better understood by analyzing the $\chi_3$ denominator $b$ (see Eq. \ref{eq_bat}) in Fig. \ref{fig1a}(b). In contrast to the case with $\Delta=0$, where $b=\lambda_\textnormal{AT}$, $b$ is always positive when $\Delta>0$. In this particular, $b$ presents a smooth minimum around the temperature $T^{*}$, leading to the rounding of $\chi_3$, whereas $\lambda_\textnormal{AT}$ becomes zero at a lower temperature. Therefore, different from the $\Delta=0$ result, $ T^*$ can identify a crossover between PM phases. This unexpected behavior is associated with RF effects.

These results for the behavior of $T_f$, $T^*$ and $T^{**}$ can be better explored in the phase diagram of Fig. \ref{fig2a}.
First of all, there are two distinct regions: one with stable RS solution ($T>T_f$) and another with RSB ($T<T_f$). 
More important, $T^*$ and $T^{**}$ are located within the RS regime, 
in which PM phases occur with different characteristics. 
For higher temperatures ($T>T^{**}$), the PM phase follows the Curie-Weiss law, in which the reciprocal of the linear susceptibility presented in the inset of Fig. \ref{linear} shows a linear extrapolation (dashed line) from high temperatures. 
As the temperature decreases, the short-range ferromagnetic interactions become more relevant introducing local ferromagnetic correlations.
These correlations enhance the cluster magnetic moment ($\bar{q}$), without bringing a long-range order due to the absence of FE intercluster interactions \cite{preschmidt}. 
Although, this mechanism is not able to bring a phase transition,  the specific heat exhibits a maximum at $T^{**}$, where these ferromagnetic correlations turn important.
In other words, some of the cluster inner degrees of freedom are frozen around $T^{**}$ favoring the stabilization of small ferromagnetic clusters. 
Moreover, the intercluster disordered interactions act on the cluster magnetic moments that are still thermal fluctuating. 
Indeed, these fluctuations become progressively slower below $T^*$ until the RSB SG transition
at $T_f$. It means that we can find more two other kinds of PM phase: one between $T^{**}$ and $T^{*}$, and another between $T^{*}$ and $T_f$.

\section{Conclusion}\label{conc}

In this work, we have analyzed  the behavior of $C_\textnormal{m}$ and $\chi_3$
in a Ising SG model formulated in terms of spin clusters with a RF following
a  Gaussian distribution with width $\Delta$. The cluster formulation results 
in two coupled problems: (i) the intercluster one,  solved exactly at mean field level; (ii) the intracluster one, understood  as the intracluster  interaction plus the inner cluster
magnetic site structure, solved exactly.

Our main results on $C_\textnormal{m}$ and $\chi_3$ are summarized in the phase diagram in Fig. (\ref{fig2a}). 
The smooth maximum of $C_\textnormal{m}$ at $T^{**}$ and the increasingly rounded maximum of $\chi_3$ at $T^*$
lead the PM phase to be unfolded in three regions: PM$_3$ for $T>T^{**}$, PM$_2$ for $T^{**}<T<T^{*}$ and PM$_1$ for $T_{f}<T<T^{*}$. 
It should be remarked that $T_f$ and $T^{*}$ decrease whereas $T^{**}$ is weakly affected as $\Delta/J$ enhances. That behavior is responsible by the enlargement of the PM$_2$ region. 
We also remark two aspects: (i) the rounded maximum of $\chi_3$ is no longer related with the onset of SG state; (ii) the nontrivial broken ergodicity corresponding to the onset of RSB SG state at $T_f$ is still given by the AT line ($\lambda_\textnormal{AT}=0$) for any value of $\Delta/J$.

The unfolding of the PM phase suggests that spin correlations develop in three stages as the temperature is lowered from the conventional high temperature paramagnetism (called here PM$_3$) until the RSB SG transition at $T_f$.  
In absence of RF, we choose $T_f\simeq 1.3 T^{**}$ by adjusting  the ferromagnetic intracluster interaction.
The smooth maximum of $C_\textnormal{m}$ at $T^{**}$ indicates that the intracluster ferromagnetic interaction starts to overcome thermal fluctuations selecting magnetic global states of the cluster, which form small 
ordered ferromagnetic regions. 
This development is illustrated by the behavior of the intensity of the cluster magnetic moment $\bar{q}$ as the temperature is lowered (see Fig. (\ref{fig_op})).  Indeed, the growth of $\bar{q}$ towards its maximum value favors the  nontrivial broken ergoditicy at $T_f$ for $\Delta=0$. 
However, when $\Delta/J$ enhances, $T^{**}$ and $T_f$ strongly deviate from the relationship $T^{**}\simeq1.3  T_f$. 
Thereby, the small ferromagnetic spin clusters formation becomes increasingly far above $T_f$. 
In particular,  the mentioned deviation is mainly caused by the behavior of $T_f$ which is quite affected by the RF. That is not the case for $T^{**}$, at least, for the range of $\Delta/J$ used in our calculations ($0\leq \Delta/J \leq 0.5$). 
In addition, as the  divergence in $\chi_3$ becomes rounded
at $T^*$, there is the onset of the PM$_1$ region. This temperature is also affected by the RF. 
Remarkably, although the RF couples with individual spins, our results show that it is the small ferromagnetic spin clusters
which play the important role to determine the rounded maximum of $\chi_3$ and the RSB SG instability. In fact, $\bar{q}$ already has its maximum value at $T_f$ and, mostly important, very close to its maximum value at $T^{*}$. 
This particular point suggests that the spin dynamics in the PM$_1$ region is  rather non-trivial. Quite probably,  a very slow one. Thus,  one can expect that in PM$_1$ region  there is a slow growth of free-energy barriers 
before the nontrivial broken ergoditicy at $T_f$. In that sense, $T^{*}$ would be a crossover temperature between two types of spin dynamics.

Since we provide an unified description of the $C_\textnormal{m}$ and $\chi_3$ 
with a RF,
we  believe that some results discussed above can have relevance for the LiHo$_x$Y$_{1-x}$F$_4$ compound  when the applied transverse field $H_t$ is weak  ($H_t< 0.5 T$)  assuming that the RF is induced by $H_t$ \cite{Laflorencie2006,Gingras2006}. 
 Our proposal is that in the LiHo$_x$Y$_{1-x}$F$_4$,  the paramagnetic phase is unfolded in three regions with  one of them (the PM$_2$ region) being a region dominated by spin short range correlations favoring clusters formation and other of them (the PM$_1$ region) acting as precursor of RSB SG that appears at lower temperatures.
 We remark that our results reproduce qualitatively not only the replacement of the  divergence in $\chi_3$ by a rounded maximum located at $T^{*}$  but also the decreasing of $T^{*}$ as $H_t$ is increased, which are observed  for small Ho concentration in {\bf the LiHo$_x$Y$_{1-x}$F$_4$ } compound 
for $H_t< 0.2 T$ (see Ref. \cite{jonsson2007}). Besides the existence of RSB SG state at lower temperature, our results also show a presence of a broad maximum located at $T^{**}$ as usually observed in SG systems.  However, it is also predicted an increasingly deviation of the relationship 
between the freezing temperature $T_f$  and  $T^{**}$,   as well
  between $T^{*}$ and $T^{**}$ as $H_t$ is increased.  This  deviation between 
$T^{*}$ and $T^{**}$ could signalize the interplay between the RF induced by $H_t$  and  small ferromagnetic spin clusters as discussed in the present work.
In fact, the deviation of the conventional relationship between $T^{**}$ and $T_f$  has been observed for  
$x=0.018$, $0.045$ and $0.08$, but, in absence of $H_t$ \cite{Quilliam07}. 
Lastly, the nature of PM$_1$ region as described above may suggest 
a Griffiths phase as precursor of the RSB SG state as proposed by Biltmo and Henelius for LiHo$_x$Y$_{1-x}$F$_4$ \cite{Biltmo2012}.
One interesting question is how robust our results are in the quantum limit, i. e., for strong  $H_t$. This limit is currently been analyzed by us.

\acknowledgments
This study was financed in part by the Coordena\c{c}\~ao de Aperfei\c{c}oamento de Pessoal de N\'ivel Superior - Brazil (CAPES) - Finance Code 001, and Conselho Nacional de Desenvolvimento Cient\'ifico e Tecnol\'ogico (CNPq).

\appendix
\section{}\label{appendixa}

The 1S-RSB parameters $q_0$, $q_1$, $\bar{q}$ and $x$ are obtained by extremizing the free energy (\ref{freeenergy}):
\begin{equation}
\overline{q}=\langle\int Dz \frac{\int Dv K^{x-1}\mbox{Tr} \sigma^2 \exp{(-\beta H_\textnormal{eff}^\textnormal{1S}})}{\int Dv K^x}\rangle_{h_i}
\end{equation}
\begin{equation}
q_{0}=  \langle \int Dz [ \frac{\int Dv  K^{x-1} \mbox{Tr} \sigma \exp{(-\beta H_\textnormal{eff}^\textnormal{1S})}}{\int Dv  K^x}]^2\rangle_{h_i}
\end{equation}
\begin{equation}
q_1= \langle \int Dz  \frac{\int Dv K^{x-2} [\mbox{Tr} \sigma \exp{(-\beta H_\textnormal{eff}^\textnormal{1S})}]^{2}}{\int Dv K^x}\rangle_{h_i}
\end{equation}
and
\begin{equation}
 \frac{x^2}{4}(q_1^2-q_0^2)=
 \langle \int Dz[  \frac{\int Dv K^{x} \ln  K^x }{\beta^2 J^2\int Dv  K^x}
-\frac{\int Dv \ln K}{\beta^2 J^2}]
 \rangle_{h_i}
\end{equation}
where the $K$ and $H_\textnormal{eff}^\textnormal{1S}$ dependence on $(\{h_i\},v,z)$ is suppressed in order to brief the equations.

\section{}\label{appendixb}

The stability analysis of the RS solution follows close to de Almeida–Thouless calculation \cite{Almeida78}. However, here, the Hessian matrix have also to consider explicitly the fluctuations on the replica diagonal 
elements. 
In this case, the replicon eigenvalue is given by the correlations:
\begin{equation}\begin{split}
 \lambda_\textnormal{AT}= 1 - \beta^2 J^2 [\langle\langle \sigma^{\alpha}\sigma^{\gamma}\sigma^{\alpha}\sigma^{\gamma}\rangle\rangle_{h_i} 
 \\
 - 
 2 \langle\langle \sigma^{\alpha}\sigma^{\gamma}\sigma^{\alpha}\sigma^{\zeta}\rangle\rangle_{h_i} +\langle \langle\sigma^{\alpha}\sigma^{\gamma}\sigma^{\delta}\sigma^{\zeta}\rangle\rangle_{h_i}]
\end{split}\end{equation}
in which $\langle \langle\cdots\rangle \rangle_{h_i}\equiv \langle\frac{\mbox{Tr} \cdots\exp[-\beta H_\textnormal{eff}(\{h_i\})]}{\mbox{Tr} \exp[-\beta H_\textnormal{eff}(\{h_i\})}\rangle_{h_i}$ and the labels $\{\alpha,\gamma,\delta,\zeta\}$ are replica indices.
In particular, these replica spin correlations result in 
\begin{equation}
 \lambda_\textnormal{AT}=1 -J^2\beta^2 \left\langle\int Dz (\langle\sigma\sigma\rangle_{H_\textnormal{eff}^\textnormal{RS}}-\langle\sigma\rangle_{H_\textnormal{eff}^\textnormal{RS}}^2
 )^2 \right\rangle_{h_i} ,
 \label{at}
\end{equation}
where $\langle \cdots\rangle_{H_\textnormal{eff}^\textnormal{RS}}=\mbox{Tr} \cdots \exp[-\beta H_\textnormal{eff}^\textnormal{RS}]/\mbox{Tr}\exp[-\beta H_\textnormal{eff}^\textnormal{RS}]$ with $H_\textnormal{eff}^\textnormal{RS}$ defined in Eq. (\ref{hrs}).

\section{}\label{appendixc}

In order to obtain $\chi_3$ within the RS solution,
the explicitly dependence of $q$ and $\bar{q}$ on $h$ have to be considered in Eqs (\ref{C1})-(\ref{C2}). We expand $q$ and $\bar{q}$ up to second order in $h$: $q= q_0 + q_2 h^2 + O(h^4)$ and $\bar{q}= \bar{q}_0 + \bar{q}_2 h^2 + O(h^4)$. 
This derivation is a tedious but straightforward calculation that results in
\begin{equation}
\chi_3=\frac{\beta^3J^2}{3} [(\frac{1}{\beta^2 J^{2}}+V_3) (\bar{q}_2-q_2) + q_2 V_2]
\end{equation}
where
\begin{equation}
\bar{q}_2- q_2=\frac{\beta^2 +\beta^2 J^2}{2-\beta^2 J^2 V_3} V_2,
\end{equation}   
\begin{equation}
 q_2=\frac{\beta^2 V_1}{2-\beta^2 J^2 (V_1+V_3)},
\end{equation}  
with
\begin{equation}
V_1=0.5(2-\beta^2 J^2 V_3)(V_3-V_2)+ \beta^2 J^2 V_2 V_4 
\label{V1}
\end{equation} 
$V_2$, $V_3$ and $V_4$ given, respectively, in Eqs. (\ref{V2}), (\ref{V3}) and  (\ref{V4}).

The expression for $\chi_3$ can be written in terms of $V_1$, $V_2$ and $V_3$ as:
\begin{equation}
 \chi_3=\frac{\beta^3}{3} \frac{1+ \beta^2 J^2 (V_1+V_3)}{[2-\beta^2 J^2 (V_1+V_3)]}  V_2.
\label{C8}
\end{equation}

\end{document}